\documentclass[aps,prd,twocolumn,superscriptaddress]{revtex4}
\usepackage{color}

\usepackage{graphicx}
\usepackage{amsmath,amssymb,latexsym}
\usepackage{bm}
\usepackage{epsfig}

\begin{document}

\title{Update on tests of the Cen A neutron-emission model of highest energy cosmic rays}

\author{Luis A.~Anchordoqui}
\affiliation{Department of Physics, University of Wisconsin-Milwaukee,
P.O. Box 413, Milwaukee, WI 53201, USA
}

\author{Haim  Goldberg}
\affiliation{Department of Physics,
Northeastern University, Boston, MA 02115, USA
}

\author{Thomas J. Weiler}
\affiliation{Department of Physics and Astronomy,
Vanderbilt University, Nashville, TN 37235, USA
}

\begin{abstract}
  \noindent
  We propose that neutron emission from Cen A dominates the cosmic ray
  sky at the high end of the spectrum. Neutrons that decay generate
  proton diffusion fronts, whereas those that survive decay produce an
  angular spike in the direction of the source. We use recent data
  reported by the Pierre Auger Collaboration to normalize the
  injection spectrum and estimate the required luminosity in cosmic
  rays. We find that such a luminosity, $L_{\rm CR} \sim 5 \times
  10^{40}~{\rm erg/s},$ is comfortably smaller than the bolometric
  luminosity of Cen A, $L_{\rm bol} \sim 10^{43}~{\rm erg/s}$. We
  compute the incoming current flux density as viewed by an observer
  on Earth, and we show that the anisotropy amplitude is in agreement
  with data at the $1\sigma$ level. Regardless of the underlying
  source model, our results indicate that after a decade of data
  taking the Pierre Auger Observatory will be able to test our
  proposal.
\end{abstract}
\pacs{xxxx}
\maketitle

About a decade ago we put forward the idea of using neutrons as
markers of ultrahigh energy cosmic ray (UHECR) emission from Cen
A~\cite{Anchordoqui:2001nt}.  In this Brief Report we update our
proposal to accommodate recent observations.

The HiRes Collaboration has reported a suppression of the CR flux
above $E_{\rm s}^{\rm HR} = 56 \pm 5 ({\rm stat}) \pm 9 ({\rm syst})
~{\rm EeV}$~\cite{Abbasi:2007sv}. The spectral index $\gamma$ of the
flux, $J \propto E^{-\gamma}$, steepens from $2.81\pm 0.03$ to $5.1
\pm 0.7$. This suppression has been confirmed by the Pierre Auger
Collaboration, measuring $\gamma = 2.69 \pm 0.2 ({\rm stat}) \pm 0.06
({\rm syst})$ and $\gamma = 4.2 \pm 0.4 ({\rm stat}) \pm 0.06 ({\rm
  syst})$ below and above $E_{\rm s}^{\rm A} = 40~{\rm EeV}$,
respectively (the systematic uncertainty in the energy determination
is estimated as 22\%)~\cite{Abraham:2008ru}.  In addition, an
intriguing CR excess was found in the direction towards Cen A, a
powerful radiogalaxy at $d= 3.4~{\rm Mpc}$~\cite{:2010zzj}.  
Out of the 69 Auger events with $E_1> 55~{\rm
  EeV}$ (collected over 6~yr but equivalent to 2.9~yr of the nominal
exposure/yr of the full Auger), the
overdensity with largest significance is 13 CRs
within $18^\circ$ from Cen A, versus only 3.2 expected
if the flux were isotropic.
Furthermore, 2 of the events arrived within less than $3^\circ$ of
the radiogalaxy.  Within errors, we set $E_{\rm s}^{\rm HR}
\simeq E_{\rm s}^{\rm A} \simeq E_1$, and take the spectrum to fall 
approximately as $J \propto E^{-4}$ above the onset of the suppression.

HiRes has presented evidence that the CR composition remains protons
up to the highest energies~\cite{Abbasi:2009nf}.  Auger data on the
depth of shower maximum $X_{\rm max}$, its rms fluctuation
$\sigma(X_{\rm max})$, and muon rates at ground level favor a heavy
composition like $^{56}$Fe~\cite{Abraham:2010yv}. A critical
assumption of our hypothesis is that the primaries of the highest
energy Auger data are protons (and neutrons) and not $^{56}$Fe.  We
note that some arguments have appeared recently which lend support to
the hypothesis of proton primaries.  One is the
reasoning~\cite{Lemoine:2009pw} that if particles responsible for the
Cen A excess are heavy nuclei (say $^{56}$Fe), then a similar (as yet
unobserved~\cite{Abreu:2011vm}) anisotropy with better statistics
should be present among protons at $E_1/26\sim 2$~EeV energies.  A
second is that modeling reveals that $X_{\rm max}$ and $\sigma(X_{\rm
  max})$ in themselves may be poor estimators of primary
composition~\cite{Ostapchenko:2010ie}.  In~\cite{Block:2007rq} it is
shown that the ratio ($k\equiv \Lambda/\lambda$) of the measured
shower attenuation length ($\Lambda$) to the interaction length of
protons in the atmosphere ($\lambda$) is likely poorly estimated in
shower codes, and poorly assigned in various CR experiments.  The
assumption of a single value~$1.26\pm0.03$~\cite{Block:2007rq} for
$k$, and proton primaries, is shown to give excellent agreement with
all CR data as of 2007.  The rate and fluctuations of early shower
development (related to depth of first interaction $X_1$, inelasticity
$K$, mean multiplicity $\langle n\rangle$, and depth of most inelastic
interaction $X_n$) are the physics behind the deviation of $k$ from
unity.  In~\cite{Wilk:2010iz} it is shown that the simple combination
$X_{\rm max}-\sigma(X_{\rm max})$ is a superior estimator of
composition than $X_{\rm max}$ or $\sigma(X_{\rm max})$
individually. The combination is much less sensitive to data features
that can skew interpretation, such as fluctuations in $X_1$ and tails
in the $X_{\rm max}$~distribution.  Furthermore, when this new
estimator is applied to the HiRes and Auger data, both data sets find
agreement with a primary spectrum of protons~\cite{Wilk:2010iz}.
Thus, we feel that the assumption herein that UHECR primaries hitting
the Earth's atmosphere are protons and neutrons is viable.

We propose that neutron emission from Cen A dominates the observed CR
flux above the suppression~\cite{Anchordoqui:2001nt}. (The
acceleration process of the parent proton population is discussed in
Appendix~\ref{ApA}.) Neutrons that decay generate the proton diffusion
fronts. The Bohm diffusive regime is usually described by an
energy-dependent diffusion coefficient, $D = 0.1~(E_{\rm EeV}/B_{ {\rm
    nG}})~({\rm Mpc}^2/{\rm Myr})$. Charged particles with $E \agt E_c
= d_{\rm Mpc} \, B_{\rm nG}~{\rm EeV}$ propagate along a straight
lines, whereas particles with $E \alt E_c$ diffuse~\cite{Sigl:1998dd}.
For an extragalactic magnetic field $B_{\rm nG} = 50$ (see
Appendix~\ref{ApB}), protons even up to the highest observed energies
undergo Bohm diffusion.  As a result of this diffusion, an injection
spectrum of $dN_0/dEdt\propto E^{-3}$ in the region of the source
cutoff, results in a spectrum $J\propto E^{-4}$ for the protons at
Earth.  The rate at Earth for the surviving neutrons is
\begin{equation}
\frac{dN_n}{dt} = \frac{S}{4 \pi d^2} \int_{E_1}^{E_2} e^{-d/\lambda(E)} \  \frac{dN_0}{dEdt} \, dE, 
\end{equation}
where $\lambda (E)\sim (E/10^{20}{\rm eV})$~Mpc is the neutron decay
length and $S$ is the area of the surface detector ($=3000~{\rm km}^2$
for Auger).  For the energy interval between $E_1 = 55~{\rm EeV}$ and
$E_2 = 150~{\rm EeV}$, we calculate the normalization factor using the
observation of 2 neutrons in 3~yr.  We then use this normalization
factor to calculate the luminosity of the source in the above energy
interval. We find $L_{\rm CR}^{(E_1, E_2)} = 9 \times 10^{39}~ {\rm
  erg/s} .$ Next, we assume continuity of the spectrum at $E_1$ as it
flattens at lower energy to $E^{-2}.$ Taking the lower bound on the
energy to be $E_0 = 1~{\rm EeV}$, we can then fix the luminosity for
this interval and find $L_{\rm CR}^{(E_0 , E_1)} = 4 \times
10^{40}~{\rm erg/s}.$ Adding these, we find the (quasi) bolometric
luminosity to be $L_{\rm CR}^{(E_0, E_2)} = 5 \times 10^{40}~{\rm
  erg/s},$ which is about a factor of 2 smaller than the observed
luminosity in $\gamma$-rays, $L_\gamma \approx 10^{41}~{\rm
  erg/s}$~\cite{Sreekumar:1999xw} in the interval $100~{\rm MeV} < E <
10~{\rm GeV}$.

To further constrain the parameters of the model, we evaluate the
energy-weighted approximately isotropic (diffuse) proton flux at
70~EeV.  The value of this diffuse flux depends on the energy
threshold $E_1$ above which Cen A is the dominate source
(55~EeV~here), on the spectral index at the source ($-3$~here), on the
nature of diffusion (Bohm~here), and on the on-time $T_{\rm on}$ of
the source~\cite{Anchordoqui:2001nt}.  We obtain $\langle E^4 \, J
\rangle = 1.6 \times 10^{57}~{\rm eV}^3 \, {\rm km}^{-2}\, \rm
{yr}^{-1} \, {\rm sr}^{-1}$, in agreement with
observations~\cite{Abraham:2008ru}, for a source actively emitting
UHECRs over $T_{\rm on}\agt 70$~Myr.  (We note that the minimum
$T_{\rm on}$ is similar to the diffusion time $T_D(E)\sim d^2/[4 D(E)]
\sim 20$~Myr at $E=70$~EeV and $B_{\rm nG} = 50$.)  If we assume
circular pixel sizes with $3^\circ$ radii, the neutrons will be
collected in a pixel representing a solid angle $\Delta \Omega \simeq
8.6 \times 10^{-3}~{\rm sr}$. The proton rate coming from the
direction of Cen A is
\begin{equation}
\frac{dN_p}{dt} = S \ \Delta \Omega \, \int_{E_1}^{E_2} \langle E^4 \, J \rangle \ \frac{dE}{E^4} = 0.08~{\rm events/yr} \, .
\end{equation}
It is important to stress that the $3^\circ$ window does not have an
underlying theoretical motivation. Recall that this angular range
resulted from a scan of parameters to maximize Auger's signal
significance. Cen A covers an elliptical region spanning about
$10^\circ$ along the major axis. Therefore, some care is required to
select the region of the sky which is most likely to maximize the
signal-to-noise~\cite{ADGW}.  

We now address the question of anisotropy. Duplicating the analysis
of~\cite{Anchordoqui:2001nt} for Bohm diffusion on a 50~nG $B$-field
we determine the amplitude of a dipole term aligned with Cen A, the
so-called ``asymmetry parameter'' $\alpha$.  For $E=70~ {\rm EeV}$ we
obtain $\alpha = 0.29$, within $1\sigma$ of the same anisotropy
amplitude $\alpha = 0.25 \pm 0.18 $ obtained from the arrival
directions of the 69 observed events~\cite{Abreu:2011ve}.

One caveat is that we assumed that neutrons completely dominate the ultrahigh energy Cen A emission spectrum; that is 
\begin{equation}
\frac{dN_0}{dE\, dt} = (N_0^n + N_0^p) E^{-3} \,, {\quad} {\rm with} {\quad}  N_0^p/N_0^n \ll 1 \, .
\label{npratio}
\end{equation}
This reduces the number of free parameters in the model. The actual proton-to-neutron fraction depends on the properties of the source, 
especially the ratio of photon-to-magnetic energy density. 

 In summary, existing data is consistent with the hypothesis that Cen A dominates the CR sky at the high end of the spectrum~\cite{Farrar:2000nw}. The Pierre Auger Observatory is in a gifted position to explore Cen A and will provide in the next 9~yr of operation sufficient statistics to test this hypothesis. We expect about 6  additional direct neutron events against an almost negligible background~\cite{note}. 

The potential detection of neutrons at Auger can subsequently be validated by the larger aperture of space-based UHECR experiments.
The JEM-EUSO mission is scheduled to launch in 2017, and remain operational aboard the International Space Station for 3 or 5 years.  Including the 20\% duty cycle and 50\% Southern versus Northern hemisphere exposure, one expects 60-100 events within $18^\circ$ of Cen A for the three-year JEM-EUSO mission, and therefore 9-15 direct neutrons. For the five-year mission, numbers are proportionately higher.  Work on the implications of our Cen A model for the JEM-EUSO mission is in progress~\cite{ADGW}.  

\section*{Acknowledgments}
We thank Soebur Razzaque and Eli Waxman for discussions.
L.A.A.\ is supported by the U.S.  National Science Foundation (NSF)  under Grant PHY-0757598 and CAREER  Award PHY-1053663. H.G.\ is supported by NSF Grant PHY-0757959. T.J.W.\ is supported by U.S. Department of Energy (DoE) Grant DE-FG05-85ER40226.

\appendix

\section{UHECR emission from Cen A}
\label{ApA}

Cen A is a complex radio-loud source identified at optical frequencies with the galaxy NGC 5128~\cite{Israel}.  Radio observations at different wavelengths have revealed a rather complex morphology. It comprises a compact core, a  jet (with subluminal proper motion $\beta_{\rm jet} \sim 0.5$~\cite{Hardcastle:2003ye}) also visible at $X$-ray frequencies, a weak counter-jet, two inner lobes, a kpc-scale middle lobe, and two giant outer lobes. The jet would be responsible for the formation of the northern inner and middle lobes when interacting with the interstellar and intergalactic media, respectively. 

In order to ascertain the capability of Cen A to accelerate UHECR
protons one first applies the Hillas criterion~\cite{Hillas:1985is}
for localizing the Fermi engine in space, {\em i.e} that the Larmor
radius be less than the size of the magnetic region.  From this
condition we infer a maximum CR energy of
\begin{equation}
\label{EMAX}
E \simeq   B_{\mu {\rm G}}\, R_{\rm kpc}~{\rm EeV} \, .
\end{equation}
Furthermore, since the magnetic field carries with it an energy
density $B^2/8\pi$ and the flow carries with it an energy flux $>
\beta \, c \, B^2 /(8\pi)$, (\ref{EMAX}) also sets a lower limit on
the rate
\begin{equation}
L_{B} > \frac{1}{8} \, \beta_{\rm jet} c R_{\rm kpc}^2 B_{\mu {\rm G}}^2  
\end{equation} 
at which the energy is carried by the out-flowing plasma, and which
must be provided by the source~\cite{Waxman:2005id}.  The minimum
total power of the jets inflating the giant lobes of Cen A is
estimated to be $\approx 8 \times 10^{43}~{\rm
  erg/s}$~\cite{Dermer:2008cy}.  This argument provides a conservative
upper limit for the magnetic field in the jet with kpc-scale radius,
$B_{\mu {\rm G}} \alt 50$, and through (\ref{EMAX}) leads to $E =
50~{\rm EeV}$~\cite{Honda:2009xd}.

Of particular interest here, it was recently noted that shear
acceleration~\cite{Rieger:2004jz} could help push proton energies up
to and beyond 50~EeV~\cite{Rieger:2009pm}. The limb-brightening in the
$X$-ray jet together with the longitudinal magnetic field polarization
in the large scale jet might be indicative of internal jet
stratification, {\em i.e.} a fast spine surrounded by slower moving
layers. Energetic particles scattered across such a shear flow can
sample the kinetic difference in the flow and will naturally
experience an additional increase in energy.  It is therefore of
interest to explore the extreme case, in which protons that diffuse
from the inner shock region into the outer shear layers
charge-exchange to produce neutrons with $E_2 \sim 150~{\rm EeV}$.
(We let experiment be the arbiter of this proposed mechanism.)

\section{Extragalactic Magnetic Field}
\label{ApB}

Surprisingly little is actually known about the extragalactic magnetic
field strength. There are some measurements of diffuse radio emission
from the bridge area between the Coma and Abell
superclusters~\cite{Kim}, which under assumptions of equipartition
allows an estimate of ${\cal O}(0.2-0.6)\,\mu$G for the magnetic field
in this region. Fields of ${\cal O}(\mu{\rm G})$ are also indicated in
a more extensive study of 16 low redshift clusters~\cite{Clarke}. It
is assumed that the observed $B$-fields result from the amplification
of much weaker seed fields. However, the nature of the initial week
seed fields is largely unknown. There are two broad classes of models
for seed fields: cosmological models, in which the seed fields are
produced in the early universe, and astrophysical models, in which the
seed fields are generated by motions of the plasma in (proto)galaxies.
Galactic winds are an example of the latter.  If most galaxies lived
through an active phase in their history, magnetized outflows from
their jets and winds would efficiently pollute the extragalactic
medium. The resulting $B$-field is expected to be randomly oriented
within cells of sizes below the mean separation between galaxies,
$\lambda_B \lesssim 1~{\rm Mpc}.$

Extremely weak unamplified extragalactic magnetic fields have escaped
detection up to now. Measurements of the Faraday rotation in the
linearly polarized radio emission from distant
quasars~\cite{Kronberg:1993vk} and/or distortions of the spectrum and
polarization properties in the cosmic microwave background
(CMB)~\cite{Barrow:1997mj,Jedamzik:1999bm} imply upper limits on the
extragalactic magnetic field strength as a function of the reversal
scale.  It is important to stress that Faraday rotation measurements
(RM) sample extragalactic magnetic fields of any origin (out to quasar
distances), while the CMB analyses set limits {\em only} on primordial
magnetic fields. The RM bounds depend significantly on assumptions
about the electron density profile as a function of the redshift.
When electron densities follow that of the Lyman-$\alpha$ forest, the
average magnitude of the magnetic field receives an upper limit of $B
\sim 10^{-9}$~G for reversals on the scale of the horizon, and $B \sim
10^{-8}$~G for reversal scales on the order of
1~Mpc~\cite{Blasi:1999hu}.  As a statistical average over the sky, an
all pervading extragalactic magnetic field is constrained to
be~\cite{Farrar:1999bw} \begin{equation} B \lesssim 3 \times 10^{-7}
  \, \left(\frac{\Omega_bh^2}{0.02} \right)^{-1} \,
  \left(\frac{h}{0.72}\right) \, \left(\frac{\lambda_B}{{\rm
        Mpc}}\right)^{1/2}~{\rm G} \, ,
\end{equation}
where $\Omega_b h^2 \simeq 0.02$ is the baryon density and $h \simeq
0.72$ is the present day Hubble expansion rate in units of
100~km/sec/Mpc. (This is a somewhat conservative bound because
$\Omega_b$ has contributions from neutrons as well as from protons,
but only electrons in ionized gas are relevant to Faraday rotation.)

Very recently it was pointed out that the study of the
energy-energy-correlation (EEC) in a given sample of UHECRs can
provide important clues on the all pervading extragalactic magnetic
field~\cite{Erdmann:2009ue}.  Since it is expected that UHECRs are
accelerated at discrete sources, the deflection in cosmic magnetic
fields would result in an energy ordering in the distribution of
arrival directions. The Pierre Auger Collaboration has recently
released their analysis of EEC~\cite{:2011pf}. The measured EEC
distribution is compatible with the expectation from isotropic arrival
directions, {\em i.e.} no energy-ordered deflections are observed near
the most energetic UHECRs.  Such an uncorrelated distribution can be
caused either by a high source density for an isotropic source
distribution or by large deflections of the UHECRs in cosmic magnetic
fields. Since the nearby distribution of matter is anisotropic we
conclude that the second option is more viable. One interpretation of
Auger data on EEC uses the PARametrized Simulation Engine for Cosmic
rays (PARSEC), to obtain a lower bound of 10 nG at 95\% CL on an
all--pervading extragalactic magnetic field with coherence length
$\lambda_B = 1~{\rm Mpc}$, assuming a source density $\leq
10^{-4}~{\rm Mpc}^{-3}$~\cite{Schiffer}. This agrees with the results
of~\cite{Anchordoqui:2001bs}. Use of these numbers should be regarded
as tentative while awaiting additional data and further independent
analyses.


\begin{thebibliography}{99}


\bibitem{Anchordoqui:2001nt}
  L.~A.~Anchordoqui, H.~Goldberg and T.~J.~Weiler,
  Phys.\ Rev.\ Lett.\  {\bf 87}, 081101 (2001)
  [arXiv:astro-ph/0103043].


\bibitem{Abbasi:2007sv}
  R.~U.~Abbasi {\it et al.}  [HiRes Collaboration],
  Phys.\ Rev.\ Lett.\  {\bf 100}, 101101 (2008)
  [arXiv:astro-ph/0703099].

\bibitem{Abraham:2008ru}
  J.~Abraham {\it et al.}  [Pierre Auger Collaboration],
  Phys.\ Rev.\ Lett.\  {\bf 101}, 061101 (2008)
  [arXiv:0806.4302].



\bibitem{:2010zzj}
  P.~Abreu {\it et al.}  [Pierre Auger Collaboration],
  Astropart.\ Phys.\  {\bf 34}, 314 (2010)
  [arXiv:1009.1855].




\bibitem{Abbasi:2009nf}
  R.~U.~Abbasi {\it et al.}  [HiRes Collaboration],
  Phys.\ Rev.\ Lett.\  {\bf 104}, 161101 (2010)
  [arXiv:0910.4184 [astro-ph.HE]].


\bibitem{Abraham:2010yv}
  J.~Abraham {\it et al.}  [Pierre Auger Collaboration],
  Phys.\ Rev.\ Lett.\  {\bf 104}, 091101 (2010)
  [arXiv:1002.0699 [astro-ph.HE]].


\bibitem{Lemoine:2009pw} 
 M.~Lemoine and E.~Waxman,
  JCAP {\bf 0911}, 009 (2009)
  [arXiv:0907.1354].
  
\bibitem{Abreu:2011vm}
  P.~Abreu {\it et al.}  [Pierre Auger Collaboration],
  JCAP {\bf 1106}, 022 (2011)
  [arXiv:1106.3048 [astro-ph.HE]].

\bibitem{Ostapchenko:2010ie} Auger data on $\sigma(X_{\rm max})$ have
  been recently interpreted in terms of a 2-component ($p$ + Fe)
  composition, assuming partial abundances $f_i$ that change smoothly
  with energy: $f_p(E) = f_p (1~{\rm EeV}) [1 - {\rm log}(E/{\rm
    EeV})/1.5]$. The data favor $f_p (1~{\rm EeV})= 0.4$.
  S.~Ostapchenko,
  arXiv:1010.0137 [astro-ph.HE].
  Such a persistent rise of the iron content in the vicinity of the ankle
  ($1~{\rm EeV} < E < 3~{\rm EeV}$) seems to be in disagreement with
  Auger measurements of the $X_{\rm max}$ distribution.


\bibitem{Block:2007rq}
  M.~M.~Block,
  Phys.\ Rev.\  {\bf D76}, 111503 (2007)
  [arXiv:0705.3037 [hep-ph]];
M.~M.~Block, F.~Halzen, T.~Stanev,
  Phys.\ Rev.\  {\bf D62}, 077501 (2000)
  [hep-ph/0004232].


\bibitem{Wilk:2010iz}
  G.~Wilk and Z.~Wlodarczyk,
  J.\ Phys.\ G {\bf 38}, 085201 (2011)
  [arXiv:1006.1781 [astro-ph.HE]].


\bibitem{Sigl:1998dd}
  G.~Sigl, M.~Lemoine and P.~Biermann,
  Astropart.\ Phys.\  {\bf 10}, 141 (1999)
  [arXiv:astro-ph/9806283].


\bibitem{Sreekumar:1999xw}
P.~Sreekumar, D.~L.~Bertsch, R.~C.~Hartman, P.~L.~Nolan and D.~J.~Thompson,
Astropart.\ Phys.\  {\bf 11}, 221 (1999) 
[arXiv:astro-ph/9901277].



\bibitem{ADGW}  L. A. Anchordoqui, P. Denton, H. Goldberg, and T. J. Weiler, in preparation.





\bibitem{Abreu:2011ve} The amplitude measurement of the first harmonic
  modulation in the right-ascension distribution has been computed
  using the classical Rayleigh formalism weighted by exposure described in
  P.~Abreu {\it et al.}  [Pierre Auger Collaboration],
  Astropart.\ Phys.\  {\bf 34}, 627 (2011)
  [arXiv:1103.2721 [astro-ph.HE]].
  The declination dependence of the relative exposure has been taken
  from
  P.~Sommers,
  Astropart.\ Phys.\  {\bf 14}, 271 (2001)
  [arXiv:astro-ph/0004016].



\bibitem{Farrar:2000nw}
  G.~R.~Farrar and T.~Piran,
  arXiv:astro-ph/0010370.



\bibitem{note} Our model also predicts no directional signals from M87.











\bibitem{Israel} F. P. Israel,
Astron. Astrophys. Rev. {\bf 8}, 237 (1998).




\bibitem{Hardcastle:2003ye}
  M.~J.~Hardcastle, D.~M.~Worrall, R.~P.~Kraft, W.~R.~Forman, C.~Jones and S.~S.~Murray,
  Astrophys.\ J.\  {\bf 593}, 169 (2003)
  [arXiv:astro-ph/0304443].



\bibitem{Hillas:1985is}
  A.~M.~Hillas,
  Ann.\ Rev.\ Astron.\ Astrophys.\  {\bf 22}, 425 (1984).




\bibitem{Waxman:2005id}
  E.~Waxman,
  Phys.\ Scripta {\bf T121}, 147 (2005)
  [arXiv:astro-ph/0502159].

\bibitem{Dermer:2008cy}
  C.~D.~Dermer, S.~Razzaque, J.~D.~Finke and A.~Atoyan,
  New J.\ Phys.\  {\bf 11}, 065016 (2009)
  [arXiv:0811.1160 [astro-ph]].


\bibitem{Honda:2009xd}   The actual size of the acceleration region can be larger because of uncertainties in the angular projection of this region along the line of sight. For example, an explanation of the apparent absence of a counter-jet in Cen A via relativistic beaming suggests that the angle of the visible jet axis with respect to the line of sight is at most 36$^{\circ}$, which could lead to a doubling of the jet radius; N. Junkes, R. F. Haynes, J. I.
 Harnett, and  D. L.  Jauncey, 
Astron. Astrophys. {\bf 269}, 29 (1993) [Erratum, {\it ibid} 
{\bf 274}, 1009 (1993)]. On the other hand, equipartition arguments seem to indicate that a field strength of $100~\mu{\rm G}$ in the inner region of the jet is plausible;
  M.~Honda,
  Astrophys.\ J.\  {\bf 706}, 1517 (2009)
  [arXiv:0911.0921 [astro-ph.HE]].
The jet power required to maintain these extreme values of $B_{\mu {\rm G}}$ and $R_{\rm kpc}$ can be reached during flaring intervals, see~\cite{Dermer:2008cy}. We conclude that acceleration of UHECR protons up to somewhat beyond 100~EeV is therefore in principle possible during powerful episodes of jet activity. 



\bibitem{Rieger:2004jz}
  F.~M.~Rieger and P.~Duffy,
  Astrophys.\ J.\  {\bf 617}, 155 (2004)
  [arXiv:astro-ph/0410269].



\bibitem{Rieger:2009pm}
  F.~M.~Rieger and F.~A.~Aharonian,
  arXiv:0910.2327 [astro-ph.HE].




\bibitem{Kim} K.-T. Kim, P. P. Kronberg, G. Giovannini and T. Venturi, Nature
{\bf 341}, 720 (1989).




\bibitem{Clarke}  T. E. Clarke, P. P. Kronberg and H. B\"ohringer, 
Astrophys. J. {\bf 547}, L111 (2001).


\bibitem{Kronberg:1993vk}
P.~P.~Kronberg,
Rept.\ Prog.\ Phys.\  {\bf 57}, 325 (1994).




\bibitem{Barrow:1997mj}
  J.~D.~Barrow, P.~G.~Ferreira and J.~Silk,
  Phys.\ Rev.\ Lett.\  {\bf 78}, 3610 (1997)
  [arXiv:astro-ph/9701063].



\bibitem{Jedamzik:1999bm}
  K.~Jedamzik, V.~Katalinic and A.~V.~Olinto,
  Phys.\ Rev.\ Lett.\  {\bf 85}, 700 (2000)
  [arXiv:astro-ph/9911100].




\bibitem{Blasi:1999hu}
P.~Blasi, S.~Burles and A.~V.~Olinto,
Astrophys.\ J.\  {\bf 514}, L79 (1999)
[arXiv:astro-ph/9812487].







\bibitem{Farrar:1999bw}
G.~R.~Farrar and T.~Piran,
Phys.\ Rev.\ Lett.\  {\bf 84}, 3527 (2000)
[arXiv:astro-ph/9906431].

\bibitem{Erdmann:2009ue}
  M.~Erdmann and P.~Schiffer,
  Astropart.\ Phys.\  {\bf 33}, 201 (2010)
  [arXiv:0904.4888 [astro-ph.HE]].

\bibitem{:2011pf}
 P. Abreu  {\it et al.}  [Pierre Auger Collaboration],
  arXiv:1107.4805.

\bibitem{Schiffer} P. Schiffer, 
PhD Thesis, RWTH Aachen University, May 2011; {\tt http://darwin.bth.rwth-aachen.de/opus3/} {\tt volltexte/2011/3698/}

\bibitem{Anchordoqui:2001bs} 
  L.~A.~Anchordoqui and H.~Goldberg,
  Phys.\ Rev.\  D {\bf 65}, 021302 (2002)
  [arXiv:hep-ph/0106217]. 




\end{thebibliography}
\end{document}